\documentclass[aps,prb,twocolumn,floats,showpacs,superscriptaddress]{revtex4-1}
\usepackage{graphicx,epsfig}
\usepackage{times}
\usepackage{graphics,dcolumn,bm,float}
\usepackage{amssymb,amsmath,rotate,color}
\usepackage[title,titletoc,toc]{appendix}
\usepackage[pagebackref=false,colorlinks,linkcolor=blue,citecolor=blue,urlcolor=magenta]{hyperref}

\begin{document}
\unitlength 1 cm
\newcommand{\be}{\begin{equation}}
\newcommand{\ee}{\end{equation}}
\newcommand{\bearr}{\begin{eqnarray}}
\newcommand{\eearr}{\end{eqnarray}}
\newcommand{\nn}{\nonumber}
\newcommand{\la}{\langle}
\newcommand{\ra}{\rangle}
\newcommand{\cd}{c^\dagger}
\newcommand{\vd}{v^\dagger}
\newcommand{\ad}{a^\dagger}
\newcommand{\bd}{b^\dagger}
\newcommand{\tk}{{\tilde{k}}}
\newcommand{\tp}{{\tilde{p}}}
\newcommand{\tq}{{\tilde{q}}}
\newcommand{\eps}{\varepsilon}
\newcommand{\vk}{\vec k}
\newcommand{\vp}{\vec p}
\newcommand{\vq}{\vec q}
\newcommand{\vkp}{\vec {k'}}
\newcommand{\vpp}{\vec {p'}}
\newcommand{\vqp}{\vec {q'}}
\newcommand{\bk}{{\bf k}}
\newcommand{\bp}{{\bf p}}
\newcommand{\bq}{{\bf q}}
\newcommand{\br}{{\bf r}}
\newcommand{\bR}{{\bf R}}
\newcommand{\up}{\uparrow}
\newcommand{\down}{\downarrow}
\newcommand{\fns}{\footnotesize}
\newcommand{\ns}{\normalsize}
\newcommand{\cdag}{c^{\dagger}}

\newcommand{\sx}{\sigma^x}
\newcommand{\sy}{\sigma^y}
\newcommand{\sz}{\sigma^z}

\title{Exactly solvable spin chain models corresponding to BDI class of topological superconductors}

\author{S. A. Jafari}
\email{akbar.jafari@gmail.com}
\affiliation{Department of Physics, Sharif University of Technology, Tehran 11155-9161, Iran}
\affiliation{Center of excellence for Complex Systems and Condensed Matter (CSCM), Sharif University of Technology, Tehran 1458889694, Iran}
\affiliation{School of Physics, Institute for Research in Fundamental Sciences, Tehran 19395-5531, Iran}
\author{Farhad  Shahbazi}
\email{shahbazi@cc.iut.ac.ir}
\thanks{Both authors have equal contribution.}
\affiliation{Department of Physics, Isfahan University of Technology, Isfahan 84156, Iran}
\affiliation{School of Physics, Institute for Research in Fundamental Sciences, Tehran 19395-5531, Iran}

\begin{abstract}
We present an exactly solvable extension of the quantum XY chain with longer range
multi-spin interactions.
Topological phase transitions of the model are classified in terms of the number of
Majorana zero modes, $n_M$ which are in turn related to an integer winding number, $n_W$. 
The present class of exactly solvable models belong to the BDI class in the 
Altland-Zirnbauer classification of topological superconductors. 
We show that time reversal (TR) symmetry of the spin variables 
translates into a {\em sliding} particle-hole (PH) transformation in the language of 
Jordan-Wigner (JW) fermions -- a PH transformation followed by a $\pi$ shift in the 
wave vector ($\pi$PH). Presence of $\pi$PH
symmetry restricts the $n_W$ ($n_M$) of TR symmetric extensions of XY to odd (even) integers.
The $\pi$PH operator may serve in further detailed classification of topological superconductors
in higher dimensions as well.
\end{abstract}
\maketitle
The spectrum of Quantum XY model is exhausted by the emergent JW fermions~\cite{LSM,Mattis2006}.
The anisotropy of the exchange coupling generates p-wave superconducting pairing between 
spinless JW fermions leading to unpaired Majorana fermion (MF) at ends of an open 
chain~\cite{Kitaev2001}.
Adding further neighbor XY couplings in general spoils the exact solvability
because the JW transformation incorporates appropriate (non-local) phase strings in order to
fulfill anti-commutation algebra~\cite{Mattis2006}. In the context of the
Ising in a transverse field (ITF) model it was recently 
shown that adding appropriately engineered three-spin
interactions can still leave it exactly solvable~\cite{Raghu2011}.
Given that ITF and XY model are related by a duality transformation~\cite{Savit,Henkel}
we expect similar extensions to work for the XY model. 
In this letter we classify generalizations of the
XY model with arbitrary $n$-spin interactions in terms of a $\pi$PH symmetry
that is a PH transformation followed by a sign alternation in one sublattice.
We show that in presence of $\pi$PH corresponding to every MF there will be
a partner MF which will correspondingly restrict the possible winding numbers.
Let us start with the XY Hamiltonian,
\begin{align}
\label{xy.eqn}
   &H_{\rm XY} = \sum_j (J_1+\lambda_1) \sx_j \sx_{j+1} + \sum_j (J_1-\lambda_1) \sy_j \sy_{j+1},
\end{align}
to which we add a $n$-spin interaction,
\be
   H_{\rm nXY}=H_{\rm XY}+\sum_{j,a}(J_r+\eta_a \lambda_r)~\sigma^{a}_j \left(\prod_{k=1}^{r-1} \sz_{j+k}\right)  \sigma^{a}_{j+r},
   \label{manyspin.eqn}
\ee
where $a=x,y$ and $\eta_x=-\eta_y=1$. 
Here $r=n-1$ denotes the range of $n$-spin interaction. 
$J_r$ is the longer range exchange and $\lambda_r$ denotes the longer range
XY anisotropy. 
For this Hamiltonian (nXY model) the quantity $q=\prod_{\ell=1}^N \sz_\ell$
is a constant of motion.
Two possible $q=\pm 1$ values correspond to number parity of JW fermions and 
hence the above generalization is expected to give a superconducting system.
Indeed the JW transformation~\cite{Mattis2006},
\be
   \sz_j=1-2\cdag_jc_j,~~\sx_j=e^{i\phi_j} (c_j+\cdag_j),~~\sy_j=-i e^{i\phi_j} (c_j-\cdag_j),
\ee
where $\phi_j$ is the phase string defined as $\phi_j=\pi \sum_{\ell < j} \cdag_\ell c_\ell$
converts the above Hamiltonian to,
\be
   H=2\sum_{s=1,r} \sum_{j} (J_s \cdag_j c_{j+s}+\lambda_s c_jc_{j+s}+{\rm h.c.}),
   \label{JW.eqn}
\ee
where longer range exchange and anisotropy parameter $J_s,\lambda_s$ 
give rise to hopping and pairing between $s$'th neighbors, respectively.
Note the important role played by $\sz$ phases is to cancel the unwanted JW phases
which renders the nXY Hamiltonian to the quadratic form~\eqref{JW.eqn}.

For even (odd) $r$ the generalized term involves $n=r+1$ spins which will be
odd (even) under the TR. 
Let us now figure out how does the JW dictionary translate the TR operation of 
spin variables. The TR changes the 
sign of spin operators $\vec\sigma_j$. Sign reversal of $\sigma^z_\ell$
with $\ell<j$ implies that under TR the non-local phase string
is transformed as $e^{i\phi_j}\to (-1)^{j-1}e^{i\phi_j}$. Therefore the
TR of spins for JW fermions translates to,
\be
   c_j\to (-1)^{j-1}c^\dagger_j \Leftrightarrow c_k\to -c^\dagger_{\pi-k}
   \label{piPH.eqn}
\ee
which is nothing but the PH transformation followed a $\pi$ shift
in the $k$-space -- or {\em sliding} PH -- that will be denoted by $\pi$PH in this paper.
The $\pi$PH can further resolve the topological classification of the Altland-Zirnbauer (AZ)
classification of topological superconductors~\cite{AZ}
that are based on PH, TR, and sublattice 
symmetries\cite{Kitaev2009,Schnyder2008,Chiu,Beenakker2015}.
In the following we discuss in detail two prototypical cases corresponding to longer
range interaction with range $r=2,3$ involving $n=3,4$ spins, respectively.
Then we present general arguments for arbitrary $r$ and discuss
an even-odd dichotomy related to the $\pi$PH transformation.

{\it 3XY Model}. 
In this case the $k$-space representation of the JW Hamiltonian is,
\begin{align}
  H= \sum_{k} {\mathbf c}^\dagger_k {\vec h}_k\cdot\vec\tau~ {\mathbf c}_k,
  ~~~~~{\mathbf c}^\dagger_k =(\cdag_k,c_{-k})
  \label{H-sq.eqn}
\end{align}
where $\tau^a,~a=x,y,z$ stands for Pauli matrices in the Nambu space and 
${\vec h}(k) = \eps_k\hat z+\Delta_k \hat y$ with
\be
   \eps_k=J_1\cos k+J_2\cos 2k,~~~
   \Delta_k=-\lambda_1\sin k-\lambda_2\sin 2k.
   \label{epsdelta.eqn}
\ee
Note that pairing with (hopping to) a neighbor at distance $r=2$ has added a
$\lambda_2\sin 2k$ ($J_2\cos 2k$) term to the Anderson pseudovector $\vec h$ representation 
of the Hamiltonian matrix. This general feature holds for any $r$. 
The eigenvalues and eigenvectors of ${\vec h}(k)$  are given by,
\be
   E_k=\pm\sqrt{\eps_k^2+\Delta_k^2},~~
   |\psi^-_k\ra= 
   \left(\!\! \begin{array}{c}
   iu_k \\
   v_k
   \end{array} \!\!\right),
   |\psi^+_k\ra= 
   \left(\!\! \begin{array}{c}
   v_k\\
   iu_k 
   \end{array} \!\!\right),
   \label{spec.eqn}
\ee
where the coherence factors are parameterized in terms of a phase
$\phi_k=\tan^{-1}(\Delta_k/(\eps_k+E_k))$ as $u_k=\sin(\phi_k/2)$ and 
$v_k=\cos(\phi_k/2)$. The JW Hamiltonian~\eqref{H-sq.eqn} is diagonalized
in terms of Bogolons $\gamma^{\dag}_k=-i u_k c^{\dag}_k+v_k c_{-k}$. 

\begin{figure}[t]
\includegraphics[width=0.55\columnwidth]{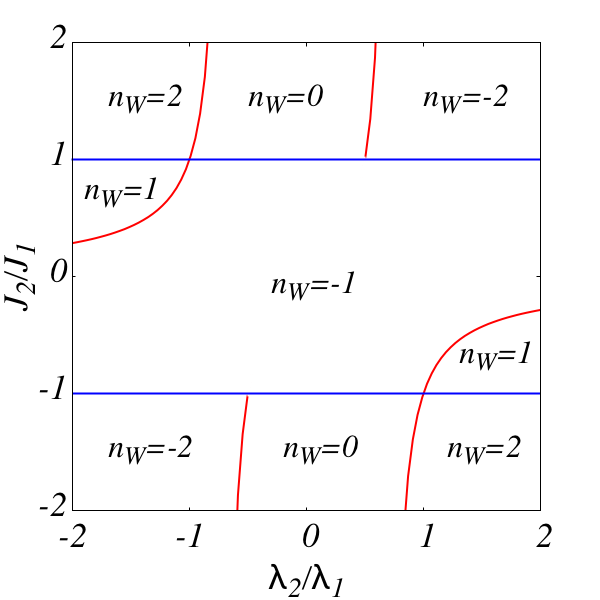}
\caption{(Color online) The phase diagram of 3XY model in parameter space.
The phase boundary curves correspond to gap closing separating topologically
distinct phases characterized by a winding number $n_W$ as indicated in the figure.
}
\label{boundary.fig}
\end{figure}

\begin{figure}[t]
\includegraphics[width=1.0\columnwidth,angle=0]{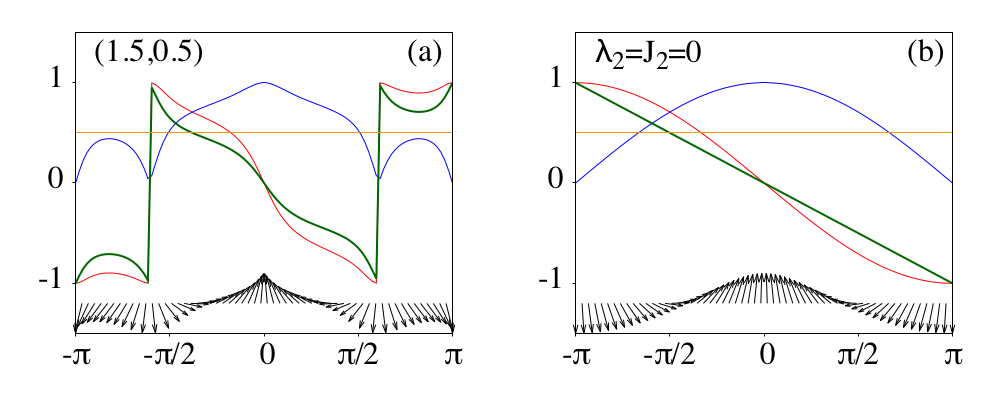}
\caption{ (Color online) Wave function and winding pattern for representative points in 
   regions various phases of Fig.~\ref{boundary.fig}.
   The red and blue curves represent the coherence factors $u_k$ and $v_k$ as a function 
   of $k$ and the green curve corresponds to the $\phi_k/\pi$. Black arrows are unit vectors 
   constructed from Anderson pseudovector $(\Delta_k,\eps_k)$ at every $k$ in the 
   1BZ. The coordinate $(\lambda_2/\lambda_1,J_2/J_1)$ is shown in the top left 
   of each panel. The resulting $n_W$ is used to label regions of Fig.~\ref{boundary.fig}.
}
\label{multiwind.fig}
\end{figure}

The boundaries of the phase diagram of the 3XY model can be analytically calculated by
investigating the gap closing of the spectrum~\eqref{spec.eqn} that happens
when both $\eps_k$ and $\Delta_k$ vanish which gives
following equations for the phase boundaries,
\begin{align}
   \frac{J_2}{J_1}=\pm 1,~~~\mbox{or}~~~
   \frac{J_2}{J_1}=\frac{(\lambda_2/\lambda_1)}{1-2(\lambda_2/\lambda_1)^2}
   ~{\rm for}~ \left|\frac{\lambda_2}{\lambda_1}\right|\geq 1/2, \label{bndry.eqn} 
\end{align}
which has been plotted in Fig.~\ref{boundary.fig}.
It is remarkable that the phase boundary 
is given in terms of ratios $J_2/J_1$ and $\lambda_2/\lambda_1$. 
This property is also general and the phase diagram is determined only in terms of 
the ratios $J_r/J_1$ and $\lambda_r/\lambda_1$.
In Fig.~\ref{boundary.fig} each region is labeled by
a winding number $n_W$. This topological invariant corresponds to the number
of times the unit circle is covered by the vector $(\Delta_k,\eps_k)$ as
$k$ varies across the first Brillouin zone (1BZ)~\cite{Alicea}. These vectors are represented by 
black arrows in Fig.~\ref{multiwind.fig} which correspond to green curve representing $\phi_k/\pi$. 
The winding pattern of arrows and the global variation profile of $\phi_k$ does not
change as long as the phase boundaries of Fig.~\ref{boundary.fig} are not crossed.
This means that $n_W$ is a topological invariant~\cite{BernevigBook}.
For the 3XY model five possible values $n_W=0,\pm 1, \pm 2$ can be extracted 
from analysis similar to Fig.~\ref{multiwind.fig}. The resulting $n_W$ values 
are used to label regions of Fig.~\ref{boundary.fig}.
The phase with $n_W=0$ is adiabatically connected to the trivially gapped phase
that can be reached by an applied field $h\to\pm\infty$ that couples to $\sz$ 
without gap closing. This makes the $n_W=0$ region topologically trivial. This is
while the other phases with $n_W\ne 0$ are separated from the trivially gapped phase
by a gap closing. The panel (b) of Fig.~\ref{multiwind.fig} corresponds to 
$\lambda_2=J_2=0$ where $\lambda_1\ne 0$ 
is adiabatically connected to the Ising limit $\lambda_1=1$. This is 
why the phase boundary~\eqref{bndry.eqn} is determined by the ratio $\lambda_2/\lambda_1$.

Let us show that for any $r$ the nXY model falls into BDI class~\cite{Schnyder2008} 
which in turn allows for winding number classification.  
First let us check the TR symmetry. For spinless fermions TR operator $\cal T$ is simply a complex
conjugation. Using the Nambu space representation, Eq.~\eqref{H-sq.eqn}, we have 
${\cal T}{\vec h}(k){\cal T}={\vec h}(-k)$
which represents the TR symmetry of the JW Hamiltonian.
Now defining the operator ${\cal C}=\tau_{x}K$ as the PH transformation,
one finds ${\cal C} {\vec h}(k) {\cal C}=-{\vec h}(k)$ that checks {\em the} PH symmetry. 
Finally for the chiral symmetry we have, 
${\vec h}(k){\cal C}{\cal T}=-{\cal C}{\cal T}{\vec h}(k)$. 
Let us emphasize 
that for every $r$, only $\sin(rk)$ functions appear in the pairing term
and hence the above properties that rest on odd parity of $\Delta_k$ apply to nXY model.
Since for the present spinless
JW fermions one has ${\cal C}^2={\cal T}^2=+1$, the nXY belongs to BDI class in the AZ
classification of topological superconductors and hence allows for 
 integer (winding number) classification of the topological phases.
However both range of possible integers and whether they are even or odd
is determined by $r$ which in turn is connected to the presence or absence
of the sliding PH symmetry, $\pi$PH. To set the stages for discussion of arbitrary $r$,
let us consider the next prototypical case of $4$-spin interactions.

\begin{figure}[t]
\includegraphics[width=0.49\columnwidth]{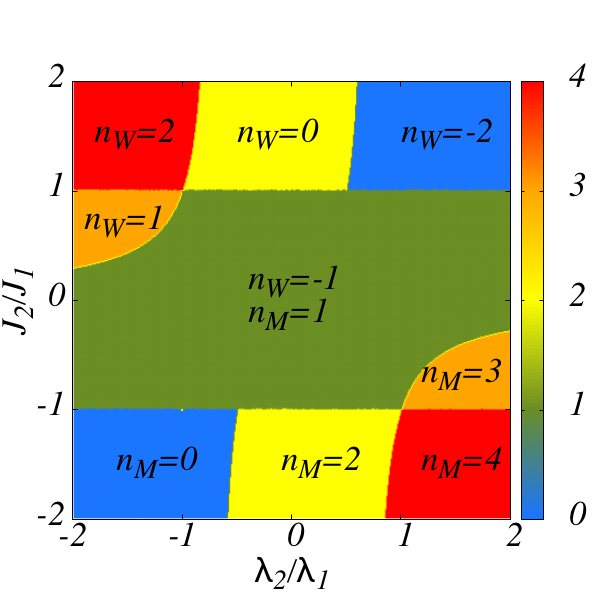}
\includegraphics[width=0.49\columnwidth]{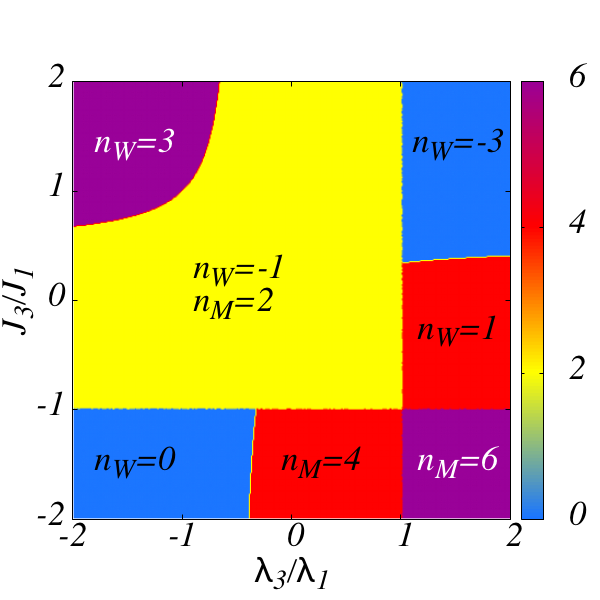}
\caption{ (Color online) Color map of the number of MFs, $n_M$ in the parameter
space for two extensions 3XY (left) and 4XY (right) corresponding to spin
clusters of range $r=2,3$, respectively. The 4XY model has $\pi$PH symmetry 
in JW representation which restricts $n_M$ to even values and $n_W$ to odd values only.
Obviously the origin in both cases correspond to $r=1$ which is not continously connected
to topologically distinct cases of $r=2$ (left) nor $r=3$ (right).
}
\label{MFcolor.fig}
\end{figure}

{\em 4XY Model}.
The four-spin term preserves the TR symmetry of the spin model. 
The components of Anderson pseudovector are given by,
$\eps_{k}=J_{1}\cos(k)+J_{2}\cos(3k)$ and $\Delta_{k}=\lambda_{1}\sin(k)+\lambda_{2}\sin(3k)$.
The phase boundaries will be given by 
setting $\eps_{k}=\Delta_{k}=0$. Therefore the gap closing curves of this model are 
the lines $\lambda_{2}/\lambda_{1}=1$ and $J_{2}/J_{1}=-1$, and the curve
\be
   \frac{\lambda_2}{\lambda_1}=1,~~\mbox{or}~~
   \frac{J_2}{J_1}=-1,~~\mbox{or}~~
   \frac{J_2}{J_1}=\frac{\lambda_{2}/\lambda_{1}}{1+2\lambda_{2}/\lambda_{1}},   
   \label{4xybound.eqn}
\ee
for  $\lambda_{2}/\lambda_{1}\geq 1$  and $\lambda_{2}/\lambda_{1}\leq -1/3$. 
These curves partition the the parameter space 
$(\lambda_2/\lambda_1,J_2/J_1)$ into seven regions represented in the right panel of 
Fig.~\ref{MFcolor.fig} each characterized by $n_{W}=\pm1, \pm 3$. The left
panel of this figure represents same set of data for 3XY model.
In both $r=2,3$ cases  $|n_W|\le r$ while for odd $r$ only odd values of $n_W$ are picked.  
To explain the meaning of the color code in this figure, let us 
discuss the number $n_M$ of MFs for a general $r$.

{\it Majorana end modes}. 
To further understand the properties of nXY model, let us now consider an open nXY
chain and discuss the Majorana zero modes of the chain.
Presence of MFs requires equal spin (or spinless pairing)~\cite{Alicea,FlensbergReview}
which engineered in one~\cite{Lutchyn} or two~\cite{FuKane} dimensions. 
In the case of XY models the spinless pairing {\em emerges}~\cite{Kitaev2001}. 
Let us now see how do MFs appear in nXY model. 
In terms of MFs $a_j = c_j+\cdag_j,~~~b_j=i(c_j-\cdag_j)$ the nXY model becomes,
\be
   H=i\sum_{s=1,r} \sum_j (J_s+\lambda_s) a_j b_{j+s} + (\lambda_s-J_s) b_j a_{j+s}.
   \label{MF.eqn}
\ee
We search for MFs of type $a$ localized
near the origin for the nXY model, i.e. zero-energy states of the form
$(A_1,0,A_2,\dots,A_N,0)$ with $A_j\sim x^j$ which gives,
\be
   (\!J_r+\lambda_r\!)+(\!J_1+\lambda_1\!)x^{r-1}\!\!-(\!\lambda_1-J_1\!)x^{r+1}\!\!
   -(\!\lambda_r-J_r\!)x^{2r}\!\!=\!0.
   \label{secul.eqn}
\ee
If we searcher for solutions of type $(0,B_1,\ldots,0,B_N)$,
we would obtain a similar equation but with $x\to x^{-1}$. 
To have a normalizable Majorana zero mode of type $a$ we need solutions
that satisfy $|x|<1$.  Fig.~\ref{MFcolor.fig} represents a color
map of the number $n_M$ of MFs of type $a$ localized in one end for $r=2,3$.
The first thing to note is that the phase boundaries in the 3XY model
obtained from the MF counting analysis precisely coincides with that in
Fig.~\ref{boundary.fig}. This is also true for the 4XY model, and the
boundaries given by Eq.~\eqref{4xybound.eqn} precisely coincide with that
in the right panel of Fig.~\ref{MFcolor.fig}. The color code in each panel
of this figure indicates the number of MFs of type $a$ bound to left end.
For each panel we have explicitly indicated $n_W$. It can observed that in both cases,
\begin{align}
   n_W=n_M-r
   \label{nWnM.eqn}
\end{align}
where $r$ is the range of interaction. Let us present a heuristic argument
that the above formula holds for any $r$.

To proceed further let us first elucidate the
meaning of $\pi$PH in the language of MFs: If we represent the JW "electron" and "hole" 
operators as $c^\dagger_j=a_j+ib_j$ and $h_j=a'_j-ib'_j$ and if we search for MFs 
with vanishing $b,b'$ component, then every zero mode solution 
$(A_1,0,A_2,0,\ldots)$ is mapped by $\pi$PH to a {\em partner} MF $(A'_1,0,A'_2,0,\ldots)$
with $A'_j=-(-1)^j A_j$. 

  For even values of $r$ where odd number of spin variables are added to the
XY model, consider any point in the phase diagram (see left panel of Fig.~\ref{MFcolor.fig}) 
with given number $n_M$ of MFs. Obviously the generalized $r+1$-spin interaction
breaks TR symmetry. In this case the overall minus arising from TR transformation can be absorbed
by the transformation $(\lambda_r,J_r)\to -(\lambda_r,J_r)$. This means that for every
MF at point $(\lambda_1,J_1,\lambda_r,J_r)$ of the phase diagram, its partner MF
corresponds to point $(\lambda_1,J_1,-\lambda_r,-J_r)$. 
This explains the inversion symmetry in the left panel of Fig.~\ref{MFcolor.fig}.
This can also be seen from Eq.~\eqref{secul.eqn} that maps to itself 
under simultaneous change of $x\to -x$ and $(\lambda_r,J_r)\to -(\lambda_r,J_r)$. 
For odd values of $r$,
there are even number of spins giving a TR symmetric term and hence the TR operation  
does not produce any minus sign in the $n$-spin term. Therefore corresponding to every MF 
at any point $(\lambda_1,J_1,\lambda_r,J_r)$, the partner MF also exists at the same
point in the parameter space. 
This can also be seen directly from Eq.~\eqref{secul.eqn}: Although 
inn general Eq.~\eqref{secul.eqn} admits $2r$ solutions such that the number
$n_M$ of them satisfying $|x|<1$ is $0\le n_M\le 2r$. However, for odd $r$ this equation
becomes an equation of degree $r$ in terms of $X=x^2$ which implies that the
solutions always come in pairs $\pm x$ giving partner MFs as $A_j\sim (\pm x)^j$.
This explains why in the right panel of Fig.~\ref{MFcolor.fig} only colors 
corresponding to even $n_M$ appear. 
The presence of $\pi$PH for odd $r$ implies that corresponding to every MF
at the chain end, its partner obtained by sign alternation in one-sublattice
is also acceptable solution and hence $n_M$ is always even.
Now let us discuss how does $\pi$PH restrict $n_W$.

Consider an arbitrary point in the phase diagram 
corresponding to pairing amplitudes $\{\lambda_s\}$.
Since the resulting JW Hamiltonian~\eqref{JW.eqn} is TR symmetric, 
the action of TR is simply $i\to -i$ which can be absorbed by 
$\{\lambda_s\}\to\{-\lambda_s\}$. The later on other hand amounts to 
changing the sing of the horizontal component $\Delta_k$ of the Anderson
pseudovector. Therefore it is the implication of TR symmetry of JW Hamiltonian 
that corresponding to every winding number $n_W$,
there is a winding number $-n_W$ irrespective of whether $r$ is even or odd. 
Now let us argue that the possible values of $n_W$ are bounded by the range 
$r$ of interaction. Every time $\Delta_k$ vanishes, the pseudovector points vertically 
either to the south or north pole. For the nXY model
the maximum number of the zeros of $\Delta_k$ in the 1BZ produced
by combination first and $r$'th harmonics of $\sin$ function 
is $r$ which implies $|n_W|\le r$.

To see when the maximum number of zeros in $\Delta_k$ are realized, it is enough
to consider the limit $J_r\sim r\lambda_r \gg \lambda_1\sim J_1$ where
there are $2r+1$ zeros for $\Delta_k$ in the 1BZ which give maximal
winding $|n_W|=r$. In this limit the secular equation~\eqref{secul.eqn}
reduces to $x^{-2r}=(r-1)/(r+1)$ the absolute value of 
which is always less than unity for every $r>0$
which realizes maximum number of MFs equal to $2r$. This means that the maximum
values of $n_W$ and $n_M$ happen in the same limit. 
Now the point with minimum $n_W$ is the TR of the maximal $n_W$. The TR for JW
Hamiltonian is equivalent to $\{\lambda_s\}\to\{-\lambda_s\}$ which is 
a $\pi/2$ rotation around the $z$-axis for spin variables and the operation  
$a\leftrightarrow b$ for MFs which essentially exchanges $x\leftrightarrow x^{-1}$
and hence mapping every state with $n_M$ MFs to a state with $2r-n_M$ MFs.
Therefore both upper and lower bounds of $2r+1$ possible integers $|n_W|\le r$ and
$0\le n_M\le 2r$ describe the same physical state. 
Finally, since both $n_M$ and $n_W$ are unique topological labels of the
same state, the mapping between the two sets must be one-to-one. Since the ends
of two chains of integers map to each other, we heuristically expect Eq.~\eqref{nWnM.eqn}
to hold for any $r$. 

Now let us discuss why for odd values of $r=2p+1$ the $n_W$ is always odd.
The $n_W$ changes by half between each two consecutive zeros of
$\Delta_k=\lambda_1\sin k +\lambda_r\sin ((2p+1)k)$. 
Suppose that this gap function
vanishes at some point $k_*$ in the 1BZ. By $\pi$PH symmetry, relation~\eqref{piPH.eqn}
it also vanishes at $\pi-k_*$. The sign of the
vertical component $\eps_k$ at $k_*$ and $\pi-k_*$ are opposite as the $\cos$ functions
appearing in vertical component $\eps_k$ of Anderson pseudovector
change sign upon going from $k_*$ to $\pi-k_*$ when $r$ is odd. 

Now starting from the XY model ($p=0$) and
focusing only in the right half of 1BZ with $k>0$, at $k=0$ and $k=\pi$ the
Anderson vectors point to north and south poles respectively
(Fig.~\ref{multiwind.fig}-b) which means that
in the right half of 1BZ one picks up a half-integer winding. 
For every $k_*$ if the winding vector points to some pole 
the one at $\pi-k_*$ will point to opposite pole, 
corresponding to every {\em pair of roots} $k_*,\pi-k_*$ of $\Delta_k$, an integer winding
is inserted to the right half of 1BZ. This means that always half-integer
windings are possible in the right half of 1BZ. Therefore the winding 
number picked over the whole 1BZ is an odd number. That is why in the 
right panel of Fig.~\ref{MFcolor.fig} we only have odd winding numbers.
The fact that for odd $r$, only even number of MFs and only odd $n_W$s
are possible is consistent with Eq.~\eqref{nWnM.eqn}. 
This line of reasoning implies that
in the simple case $p=0$ corresponding to XY (or Kitaev) chain, 
by $\pi$PH symmetry the topologically non-trivial phase always hosts
two independent Majorana end modes related by $A_j\leftrightarrow -(-1)^j A_j$.

It can be noticed that the $Z_2$ index defined by 
$\nu={\rm sign}(\eps_0){\rm sign}(\eps_\pi)$~\cite{Alicea},
gives $\nu=-1$ when $r$ is odd. However, when $r$ is even, 
$\nu={\rm sign}(|J_r/J_1|-1)$. For even $r$, the $\nu=+1(-1)$ corresponds 
to even (odd) values of $n_W$.
The physical interpretation of $\nu$ is as follows: 
When $\nu=+1 (-1)$ the identity of Bogolons does not change (changes) 
from hole-like to particle-like when $k$ spans the range $[0,\pi]$. 
The presence of $\pi$PH symmetry for odd $r$ guarantees that the above
$Z_2$ index takes only one value $-1$ which means that the charge character of
Bogolons at $k=0$ and $k=\pi$ are opposite. Breaking $\pi$PH allows for both
$\pm1$ values.




{\em Summary}. To summarize, we have presented an exactly solvable extension of the quantum
XY model that involves clusters of $n=r+1$ spins interacting at range $r$. 
The ensuing JW representation is a topological superconductor in BDI  class.
We showed that the TR operation of original spin variables translates to 
a sliding PH transformation of JW fermions, $\pi$PH. The presence of $\pi$PH
implies that corresponding to every MF wave function $A_j$, 
there is a partner MF whose wave function is $-(-1)^jA_j$ which in turn 
restricts the number of MFs to even values only. The $\pi$PH also implies
that the roots of pairing potential come in pairs which
restricts the $Z$ winding numbers to odd integers only. 
The Bott periodicity~\cite{Kitaev2009} implies that there should exist similar restriction
in higher dimensions on topological invariants when $\pi$PH is a symmetry. 
It will be interesting
to study possible higher dimensional models with $\pi$PH 
symmetry in electronic systems~\cite{FuKane,ReadGreen,Stone}. 
The number $n_M$ of MFs leaves a unique signature in tunneling experiments
and hence remains directly accessible to experiments. 
Array of magnetic nano-particles on a superconductor is described by an 
effective theory that includes $r=2$ hopping between the spinless fermions~\cite{Beenakker2015,Choy2011}
which may serve as potential platform to materialize 3XY model.

{\em Acknowledgement}.
We thank S. Moghimi Araghi and A. T. Rezakhani useful discussions.
We are grateful to C. W. J. Beenakker, Alexander Altland,  
Mehdi Kargarian and Anthony Leggett for insightful comments and communictations.

\end{document}